\documentclass[prl,twocolumn,floatfix]{revtex4}
\usepackage{amsmath}
\usepackage{graphicx}
\usepackage{bm}
\usepackage{amssymb}
\usepackage{color}

\newcommand{\be}{\begin{equation}}
\newcommand{\ee}{\end{equation}}

\def \be{\begin{equation}}
\def \ee{\end{equation}}
\def \ba{\begin{array}}
\def \ea{\end{array}}
\def \bea{\begin{eqnarray}}
\def \eea{\end{eqnarray}}

\begin{document}

\title{Fractional charges on an integer quantum Hall edge}
\author{E. Berg$^1$, Y. Oreg$^{1,2}$, E.-A. Kim,$^{1,3}$ and F. von Oppen$^4$}
\affiliation{$^{1}$ Department of Physics, Stanford University, Stanford, CA 94305-4045, USA%
\\
$^{2}$ Department of Condensed Matter Physics, Weizmann Institute
of Science,
Rehovot, 76100, Israel\\
$^{3}$ Department of Physics, Cornell University, Ithaca, NY 14853 \\
$^{4}$Institut f\"{u}r Theoretische Physik, Freie Universit\"{a}t
Berlin, Arnimallee 14, 14195 Berlin, Germany}

\begin{abstract}
We propose ways to create and detect fractionally charged excitations in
\emph{integer} quantum Hall edge states. The charge fractionalization occurs
due to the Coulomb interaction between electrons propagating on different
edge channels. The fractional charge of the soliton-like collective
excitations can be observed in time resolved or frequency dependent shot
noise measurements.
\end{abstract}

\date{\today }
\maketitle

\emph{Introduction$-$} Fractionalization in low dimensional
systems is a striking example of emergent behavior caused by
strong correlations. Well known examples include fractionally
charged excitations in one-dimensional charge density wave
systems~\cite{SuHeeger}, and in the fractional quantum Hall (FQH)
effect~\cite{fqhe}. In particular, the detection of fractional
charge by measuring shot noise in the point contact scattering
current between FQH edge states\,\cite{fqhe_exp,KF} made the
latter a celebrated example.

Pham \textit{et. al.} have predicted\thinspace \cite{Pham} that an electron
injected into an interacting wire will be fractionalized into right and left
moving excitations, each carrying a non-integer charge that depends on the
Luttinger parameter~$g$. The observation of this effect is a considerable
challenge, because it occurs inside the interacting wire, while most
measurements are made in the Fermi liquid leads. Strong evidence for
electron fractionalization has recently been given in GaAs quantum wires
\cite{Hadar,Karyn} by a clever analysis of transport measurements. However,
a direct detection of the fractional charge is desirable.

In this Letter, we propose ways to create and detect excitations
with well-defined fractional charges by injecting electrons into
integer quantum Hall (IQH) edge states. Unlike the FQH
case~\cite{fqhe_exp}, the fractional charge of these collective
excitations is not associated with fractional quasiparticles in
the bulk, but rather results from Coulomb interactions between
electrons on the edges
\cite{YuvalFinkelstein,Kim-Fradkin,Sukhorukov}. The role of the
bulk integer QH state is to provide edge states which would form a
chiral Fermi liquid~\cite{Halperin} in the absence of Coulomb
interactions. An important advantage of our IQH setting is the
spatial separation of the edge states of opposite chirality, which
allows separate access to each edge. For instance, the current can
be injected into one edge, while the backscattered current is
collected on the other.

\begin{figure*}[t]
\centering
\includegraphics[width=13.0cm]{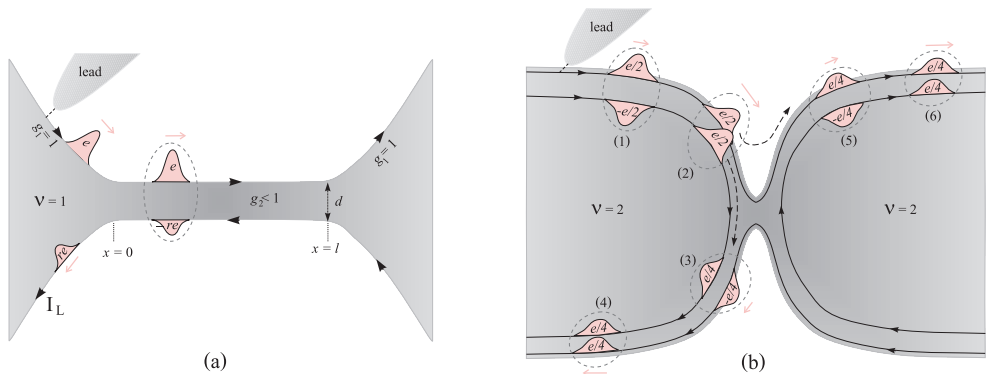}
\caption{(Color online.) Geometries of the proposed experiments.
See text for details. (a) $\protect\nu=1$ geometry. The shaded
region is an IQH bar. In the central narrow region inter-edge
interactions are significant, leading to an interaction parameter
$g<1$, while everywhere else $g=1$. (b) $\protect\nu=2$ geometry.
In the narrow region, the inner edge mode is reflected, while the
other is transmitted.} \label{fig:geometry}
\end{figure*}

We propose to detect the charge fractionalization by specific time
resolved or finite frequency
\cite{YuvalFinkelstein,Safi,Ponomarenko} shot noise experiments,
which can directly measure the charge of the elementary carriers.
To demonstrate this, we calculate explicitly the shot noise in the
two proposed geometries shown in Fig. \ref{fig:geometry}. We will
now discuss these geometries in detail.

$\nu =1$\emph{\ geometry-} (Fig. \ref{fig:geometry}a) This
geometry consists of a pair of counter propagating IQH edge
states, which are close enough for significant inter-edge
interactions in the center region, and a lead which injects
electrons into one of the edge states via tunneling. For
simplicity, we assume that the electron spins are completely
polarized along the magnetic field. The pair of edge states can be
modelled as a non-chiral Luttinger liquid (LL) with position
dependent interaction parameter $g(x)$, which
varies smoothly (on the scale of the magnetic length $\ell _{B}$) from $%
g_{1}=1$ to $g_{2}<1$, and back to $g_{1}$. The value of $g_{2}$
is determined by the strength of the inter-edge interaction in the
center region. We assume that the inter-edge separation in the
center region is large enough so that inter-edge tunneling is
negligible, while small enough to allow for significant inter-edge
interactions. This is possible in principle, since the tunneling
is suppressed exponentially with the inter-edge distance
$d$~\cite{comment-tunneling}, while the
interaction decays only as a power law. The lead is biased with voltage $V_{%
\mathrm{lead}}$, relative to the upper (right moving) edge.

Fractionalization due to interactions in the central region
manifests itself through the reflection of a fractional charge
$q^{\ast }=re$, with $r=(1-g_{2})/(1+g_{2})<1$ \cite{SafiSchulz},
in the lower edge each time an injected electron hits the $x=0$
boundary between the non-interacting and interacting regions (see
Fig. \ref{fig:geometry}). This is a consequence of the fact that
the right moving eigenmode of the interacting region consists of
electrons of \emph{both} chiralities~\cite{ML,Pham}. In this
region, the injected electron in the upper edge induces a
\textquotedblleft mirror\textquotedblright\ charge $-q^{\ast }$ on
the lower edge \cite{Comment08-smooth-transition}. Since charge is
conserved on each edge separately (due to the absence of
inter-edge tunneling), this requires a simultaneous reflection of
charge $q^{\ast }$ in the lower edge.% as the injected electron
%propagates into the interacting region.

It is important to emphasize that $r$ is \emph{not} a quantum
amplitude for electron reflection~\cite{SafiSchulz, Karyn}. A
fractional charge $q^{\ast }$ is reflected to the lower edge
\emph{each time} an electron tunnels in from the lead to the upper
edge. In fact, $r$ is the reflection coefficient of the edge
plasmon modes in the infinite wavelength limit.

The propagating mode in the interacting region is later partially reflected
from the $x=\ell $ boundary in a similar process. This repeats
%\thinspace\cite{Comment08-n-reflection}
alternately at the $x=0$ and $x=\ell $ boundaries. Eventually, the
net reflected charge in the lower edge is zero, and the net
transmitted charge on the upper edge is $e$. This follows from the
separate charge conservation laws on the two edges.% and the
%absence of interactions at large distances.

Clearly, in order to detect the excitations with a well defined charge $%
q^{\ast }$ created in the first reflection, it is necessary to
avoid later reflections that carry a different charge. This can be
done formally by sending $\ell$, the length of the interacting
wire, to infinity, hence absorbing all the transmitted charge. We
consider the noise in the reflected current at frequency $\omega$,
defined as
\begin{equation}
S_{L}\left( \omega \right) =\int_{-\infty }^{\infty }dte^{i\omega
t} \left[ \left\langle \left\{ I_{L}\left( t\right) ,I_{L}\left(
0\right)
\right\} \right\rangle -2\left\langle I_{L}\left( 0\right) \right\rangle ^{2}%
\right] \text{,}  \label{S_omega2}
\end{equation}%
where $I_{L}$ is the left moving current and $\{...\}$ represents an
anti-commutator. Taking the limit $\ell\rightarrow \infty $ first and then $%
\omega \rightarrow 0$, one expects the noise at temperature $T$ to take the
form
\begin{equation}
S_{\ell }(\ell \rightarrow \infty ,\omega \rightarrow 0)=S_{\mathrm{tun}%
}\left( \omega \rightarrow 0\right) +S_{0}\left( \omega \rightarrow 0\right)
\text{.}
\end{equation}%
Here, the noise due to the tunneling from the lead
$S_{\mathrm{tun}}\left( \omega \rightarrow 0\right)$ has the form
(assuming uncorrelated tunneling events)
\begin{equation}
S_{\mathrm{tun}}\left( \omega \rightarrow 0\right) =2q^{\ast
}\coth \left( \frac{eV_{\mathrm{lead}}}{2T}\right) \langle
I_{L}\rangle \text{,} \label{S_omega1}
\end{equation}%
which depends explicitly on $q^{\ast }=re$. $V_{\mathrm{lead}}$ is
the lead bias voltage, and $S_{0}\left( \omega \right) =
\frac{e^{2}}{2\pi }\omega\coth \left( \frac{\omega }{2T}\right) $
is the LL noise in the absence of tunneling~\cite{Chamon}. We have
set $\hbar =k_{B}=1$. The main steps in the derivation of
Eq.~(\ref{S_omega1}) will be outlined below.

In a finite size system, we propose two ways to observe the charge
fractionalization. First, the fractionalization has imprints in
the finite frequency noise of reflected current, similar to the
case of an impurity in a LL\thinspace \cite{Safi}. Second, we
propose a scheme for recovering Eq. (\ref{S_omega1}) even for a
finite system. The measurement is divided into cycles. In each
cycle, $V_{\mathrm{lead}}$ is turned on for a time interval
$\Delta T_{0}\lesssim 2\ell /u$, where $u$ is the charge velocity
in the interacting region, and then turned off. The backscattered
current and noise are then measured over a time window which
extends from $t=T_{0}$ to $t=T_{0}+\Delta T_{0}$, where $T_{0}$ is
the time interval between the tunneling of an electron from the
lead and the arrival of a reflected charge to the detector in the
lower edge. This ensures that only reflections from the $x=0$
boundary are detected. The
measurement is then stopped for a time interval of a few times $\Delta T_{0}$%
, during which the excess charge in the interacting region decays.
The measurement cycle is then repeated. The noise averaged over
many cycles should satisfy Eq. (\ref{S_omega1}), from which the
fractional charge $q^{\ast }$ can be extracted.

The above procedure requires that $\Delta T_0$ is much longer than
the characteristic time of a single tunneling event, $\Delta
T_{\mathrm{tun}}$. Since electrons are injected at energy
$eV_{\mathrm{lead}}$, $\Delta T_{\mathrm{tun}} \sim
1/eV_{\mathrm{lead}}$~\cite{neder}, leading to the additional
condition $eV_{\mathrm{lead}} \gg 1/\Delta T_0$.

We now derive Eq. (\ref%
{S_omega1}), as well as a general formula for the frequency dependent noise
in the backscattered current. The system is described by the Hamiltonian
\begin{equation}
\mathcal{H}=\mathcal{H}_{1}+\mathcal{H}_{\text{\textrm{lead}}}+\mathcal{H}_{%
\mathrm{tun}}\text{,}
\end{equation}%
where $\mathcal{H}_{1}$ is the LL Hamiltonian
\begin{equation}
\mathcal{H}_{1}=\int dx{\Big \{}v\left[ \left( \partial _{x}\phi _{R}\right)
^{2}+\left( \partial _{x}\phi _{L}\right) ^{2}\right] +2V\left( x\right)
\partial _{x}\phi _{L}\partial _{x}\phi _{R}{\Big \}}\text{.}  \label{H}
\end{equation}%
Here $\phi _{R}$, $\phi _{L}$ are bosonic fields describing the two chiral
edge modes, satisfying $\left[ \phi _{R/L}\left( x\right) ,\phi _{R/L}\left(
x^{\prime }\right) \right] =\pm \frac{i}{4}\mathrm{sgn}\left( x-x^{\prime
}\right) $, where the upper (lower) sign corresponds to the right (left)
moving field, and $\left[ \phi _{R}\left( x\right) ,\phi _{L}\left(
x^{\prime }\right) \right] =\frac{i}{4}$\textrm{. } $v=v_{F}+U$ where $v_{F}$
is the \textquotedblleft bare\textquotedblright\ (noninteracting) Fermi
velocity, and $U$, $V\left( x\right) $ are the intra-edge and inter-edge
interaction strengths, respectively. We assume, for simplicity, that $U$ is
position independent, and that $V\left( x\right) =0$ for $x<0$ and $x>\ell $%
. The Luttinger parameter and the charge velocity are given by
$g\left( x\right) =\sqrt{\frac{v-V\left( x\right) }{v+V\left(
x\right) }}$ and $u\left( x\right)  =\sqrt{v^{2}-\left[ V\left(
x\right) \right] ^{2}}$, respectively.

%$g\left( x\right) $ and the velocity $u\left( x\right) $ are
%related to the parameters in Eq. (\ref{H}) by $g\left( x\right)
%&=&\sqrt{\frac{v-V\left( x\right) }{v+V\left( x\right) }}$ and
%\begin{eqnarray}
%g\left( x\right)  &=&\sqrt{\frac{v-V\left( x\right) }{v+V\left( x\right) }}%
%\text{,\ }  \label{g} \\
%u\left( x\right)  &=&\sqrt{v^{2}-\left[ V\left( x\right) \right] ^{2}}\text{.%
%}  \label{u}
%\end{eqnarray}
$\mathcal{H}_{\text{lead}}$ is the lead Hamiltonian
\begin{equation}
\mathcal{H}_{\mathrm{lead}}=\sum_{k}(\varepsilon _{k}^{%
\vphantom{\dagger}}+eV_{\mathrm{lead}})c_{k}^{\dagger
}c_{k}^{\vphantom{\dagger}}\text{,}
\end{equation}
where $\varepsilon_k$ are the single particle levels in the lead,
and $\mathcal{H}_{\mathrm{tun}}$ is the tunneling Hamiltonian
between the lead and the upper (right moving) edge:
\begin{equation}
\mathcal{H}_{\mathrm{tun}}=-\gamma \psi _{R}^{\dagger }\left( x_{0}\right)
c^{\vphantom{\dagger}}\left( x_{0}\right) +h.c.
\end{equation}
$c_{k}^{\dagger }$ is a creation operator of an electron in the lead, and $%
\psi _{R}^{\dagger }$ is the creation operator of a right moving electron.
The tunnel junction is located at $x_{0}<0$.

In order to calculate the backscattered current $\left\langle
I_{L}\right\rangle $ and the frequency dependent noise
$S_{\mathrm{tun}}\left( \omega \right) $, we use the standard
non-equilibrium Keldysh formalism\thinspace \cite{Keldysh}. Both
the current and the noise are calculated to second order in the
tunneling amplitude $\gamma $, assuming it is small (which is
necessary to ensure Poisson distributed tunneling events). We omit
the details of this calculation, which are similar to those
of~\cite{Chamon}, and state only the results below. The
backscattered current is
\begin{equation}
\left\langle I_{L}\left( x_{1}\right) \right\rangle =r\left( \omega
\rightarrow 0\right) I_{\mathrm{tun}}\text{.}
\end{equation}%
$r\left( \omega \right) $ is the frequency-dependent reflection coefficient
\begin{equation}
r\left( \omega \right) =-2iv_{F}\int_{0}^{\infty }dte^{i\omega
t}\left\langle \left[ \partial_x \phi _{L}\left( x,t\right) ,\phi
_{R}\left( x_{0},0\right) \right] \right\rangle \text{.}
\label{r_omega}
\end{equation}%
$r(\omega )$ defined in Eq.~(\ref{r_omega}) coincides with the
reflection coefficient of the edges plasmon
modes~\cite{YuvalFinkelstein}.
$I_{\mathrm{tun}}=\frac{e^{2}\left\vert \gamma \right\vert
^{2}}{v_{F}}N(0)V_{\mathrm{lead}}$ is the tunneling
current~\cite{mahan}, where $N(0)$ is the density of states of the
lead at the Fermi energy.

In the important regime $eV_{\mathrm{lead}}\gg \omega$, the shot
noise in the reflected current takes the simple form
%[after subtracting
%contribution $S_{0}\left( \omega \right) $, which is unrelated to
%the tunneling current] is
\begin{equation}
S_{\mathrm{tun}}\left( \omega \right) =2er\left( \omega \right)
r\left( -\omega \right) \coth \left(
\frac{eV_{\mathrm{lead}}}{2T}\right) I_{\mathrm{tun}}  \label{St}
\end{equation}%
where $r\left( \omega \right) $ is given by Eq.\thinspace
(\ref{r_omega}). Here, we have subtracted the $S_0(\omega)$ term,
which is unrelated to the tunneling from the lead. We see that
$S_{\mathrm{tun}}\left( \omega \rightarrow 0\right)$ satisfies Eq.
(\ref{S_omega1}).

As we noted before, for a finite length of the interacting region,
$r\left( \omega \rightarrow 0\right) =0$ and therefore both
$\left\langle I_{L}\left( x_{1}\right) \right\rangle $ and
$S_{\mathrm{t}}\left( \omega \rightarrow 0\right)$ vanish. We
demonstrate the signatures of charge fractionalization in the
finite frequency noise by calculating the noise explicitly from Eq. (\ref{St}%
) for the case of a \textquotedblleft step\textquotedblright\
variation of the inter-edge interaction strength, \emph{i.e.}
$g\left( x\right) =g<1$ for $0<x<\ell $ and $g\left( x\right) =1$
elsewhere. In this case, the reflection coefficient can be found
analytically by considering the infinite sequence of reflections
from the two boundaries. The time dependent reflection coefficient
is \thinspace \cite{SafiSchulz}
\begin{equation}
r\left( t\right) =r_{0}\delta \left( t\right) +t_{0}r_{0}^{\prime
}t_{0}^{\prime }\sum_{n=0}^{\infty }\left( r_{0}^{\prime }\right)
^{2n}\delta \left[ t-\left( n+1\right) \Delta T\right]   \label{r_t}
\end{equation}%
where $r_{0}=\frac{1-g}{1+g}$ ($r_{0}^{\prime }=\frac{g-1}{g+1}$) and $t_{0}=%
\frac{2g}{1+g}$ ($t_{0}^{\prime }=\frac{2}{1+g}$) are the reflection and
transmission coefficients from the non-interacting to the interacting
(interacting to non-interacting) boundary, respectively, and $\Delta T=\frac{%
2\ell }{u}$. Fourier transforming Eq. (\ref{r_t}), we get
\begin{equation}
r\left( \omega \right) =r_{0}\frac{1-e^{i\omega \Delta T}}{%
1-r_{0}^{2}e^{i\omega \Delta T}}.  \label{r_w}
\end{equation}%
The resulting shot noise from Eq.(\ref{St}) is peaked at $\omega =\frac{\pi
}{\Delta T}$, and both its height and width depend on $g$.
%Note that the periodic nature of %T
%the function $r(\omega )$ in frequency (with period $\frac{2\pi }{\Delta T}$%
%) is an artifact of the abrupt transition from $g=1$ to $g<1$. A smooth
%interface extending over a region of size $l\ll L$ would suppress $r(\omega
%) $ for $\omega \gtrsim 2\pi u/l$.

In the above expression, the characteristic frequency of the noise
spectrum is $\frac{1}{\Delta T}=u/\ell $. This sets the required
time resolution for detecting charge fractionalization. Assuming
that $\ell $ is as large as a few mm and~$u\!\sim
\!10^{5}\!-\!10^{6}$m$/$s\thinspace \cite{ashoori,fujisawa}, the
above characteristic frequency is of the order of
$10^{2}$-$10^{3}$MHz. In order for Eq. (\ref{St}) to hold,
$eV_{\mathrm{lead}}\gg \omega\sim 0.01-0.1 \mu$eV is required.

We roughly estimate the typical values of the interaction
parameter $g$, and hence the reflected fractional charge $q^{\ast
}$. $g$ depends on the intra-edge interaction $U$ and inter-edge
interaction $V $. Measurements of the magnetoplasmon frequency on
a single ($V\!=\!0$) IQH edge\thinspace \cite{Talyanskii} indicate
that $U\geq v_{F}$.
%Hence we get that
%$\frac{v_{F}}{U}\leq 1$ (since the gate can at most screen the
%intra-edge interaction completely, in which case $u=v_{F}$). From
%Eq. (\ref{g}) we see that setting $v_{F}=U$ gives an upper bound
%on $g$. Next, we need to estimate $\frac{V}{U}$.
In order to estimate the value of $V/U$, we model the pair of edge states as
cylindrical wires of radius $a\sim \ell _{B}$, at a distance $d$ apart.
Assuming a screened Coulomb interaction with screening length $l_{sc}>d$, $%
U\sim \frac{e^{2}}{\varepsilon \pi }\ln \frac{l_{sc}}{a}$ and $V\sim \frac{%
e^{2}}{\varepsilon \pi }\ln \frac{l_{sc}}{d}$, where $\varepsilon $ is the
dielectric constant of the surrounding semiconductor. For a rough feeling on
the typical values of $g$, we use $l_{sc}=2d$ and $d=10a$, for which $\frac{V%
}{U}\simeq 0.3$. This gives (assuming that $U=v_{F}$, which yields an upper
bound on $g$) $g\simeq 0.86$, and therefore $q^{\ast }=\frac{1-g}{1+g}%
e\simeq 0.075e$. Due to the logarithmic dependence of $U$ and $V$ on the
geometrical parameters, $g$ is not extremely sensitive to the geometry as
long as $l_{sc}$ is large enough. Similar estimates of $g$ were obtained in
Ref.~\cite{girvin}.

\emph{$\nu =2$ geometry-} This geometry, shown in Fig.
\ref{fig:geometry}b, consists of a $\nu =2$ IQH liquid with two
chiral edge modes of opposite spin. A constriction in the middle
reflects only the inner edge mode, while the outer one is
transmitted. We assume that the single-particle inter-channel
scattering is negligible~\cite{McEuen}. The two chiral edge modes
are described by the Hamiltonian
\begin{equation}
\mathcal{H}_{2}=\int dx{\Bigg [}\sum_{i=1,2}v_{i}\left( \partial _{x}\phi
_{i}\right) ^{2}+2V\partial _{x}\phi _{1}\partial _{x}\phi _{2}{\Bigg ]}%
\text{,}  \label{H2}
\end{equation}%
where $\phi _{i}$ ($i=1,2$) are the (chiral) bosonic fields for
the outer and inner edge mode, respectively, $v_{i}=v_{F,i}+U_{i}$
where $v_{F,i}$ and $U_{i}$ are their Fermi velocity and
intra-edge mode interaction, and $V$ is the interaction between
the two modes.

As in the $\nu =1$ case, electrons tunnel into the IQH edge from a
lead. We assume that the electrons couple only to the outer
($i=1$) edge mode.
However, due to the inter-mode interaction, the eigenmodes of $\mathcal{H}%
_{2}$ are combinations of charge excitations on \emph{both} edge
modes. Therefore, the injected electron is decomposed into two
eigenmodes [indicated as (1) and (2) in Fig. \ref{fig:geometry}b].
For simplicity, let us consider the case $v_{1}=v_{2}\equiv v$. In
this case, the charges of the two eigenmodes are $Q_{\pm }=\left(
q^{\ast }\text{,}\pm q^{\ast }\right) $ where $q^{\ast
}=\frac{e}{2}$ (the two components of $Q_{\pm}$ are the charges on
the outer and inner edge modes, respectively), moving at
velocities $u_{\pm }=v\pm V$. The faster even ($+$) mode reaches
the point contact first. It then splits into two $\frac{e}{2}$
packets, one moving to the right and the other reflected to the
lower, left moving
edge. Both charge packets then split again, as indicated in Fig. \ref%
{fig:geometry}b(3-6), into even (charge $\frac{e}{2}$) and odd
(charge $0$) modes, moving at velocities $u_\pm$.

The odd ($Q_{-}$) mode reaches the point contact later, and splits into a $%
-q^{\ast }$ packet scattered to the left and a $q^{\ast }$ packet
transmitted to the right. Thus, as in the $\nu =1$ case, the
\emph{net} effect (after a sufficiently long time) is the
transmission of a single electron (charge $e$) to the right. The
intermediate charge fractionalization can be detected by measuring
either the finite frequency noise spectrum of the transmitted or
reflected currents, or by performing a time resolved measurement,
similar to the one described in the $\nu =1$ case.

The finite frequency noise spectra in the transmitted and reflected currents
$S_{r,t}\left( \omega \right) $ can be calculated very similarly to Eq. (\ref%
{St}). The result is (assuming $eV_{\mathrm{lead}} \gg \omega$)
\begin{equation}
S_{t,r}\left( \omega \right) =2e\left\vert \alpha _{t,r}\left(
\omega \right)
\right\vert ^{2}\coth \left( \frac{eV_{\mathrm{lead}}}{2T}\right) I_{\mathrm{%
tun}}\text{,}  \label{Srt}
\end{equation}%
where $I_{\mathrm{tun}}$ is the tunneling current from the lead, $\alpha
_{t}\left( \omega \right) =\cos \left( \frac{\omega \Delta \tilde{T}%
}{2}\right) $ and $\alpha _{r}\left( \omega \right) =\sin \left(
\frac{\omega \Delta \tilde{T}}{2}\right) $ are the Fourier
transformed transmission and reflection coefficients. $\Delta
\tilde{T}=\ell \left( \frac{1}{u_{-}}-\frac{1}{u_{+}}\right) $
where $\ell $ is the total length of the IQH\ edge from the lead
to the detector. %The charge fraction $q^{\ast }=\frac{e}{2}$ can
%thus be estimated by measuring the noise at frequencies of the
%order of $\frac{1}{\Delta \tilde{T}}$ and fitting to
%Eq.~(\ref{Srt}).

In the more general case where $v_{1}\neq v_{2}$, other values of
$q^{\ast }$ can be obtained. The analysis in this case is
straightforward, but slightly more involved, and will be presented
elsewhere.

\emph{Conclusions- }We propose ways to create and detect
fractional charges on chiral edges of IQH liquids. The main
advantage of using IQH\ edges for this purpose is their high
controllability. In the proposed experiments, electrons are
injected into IQH edges and \textquotedblleft
split\textquotedblright\ due to Coulomb interactions into
fractionally charged packets. In the $\nu =1$ setup, this occurs
as a result of interactions between counter-propagating edge
modes. In the $\nu =2$ case, it occurs due to interactions between
modes of the same chirality. (The latter case generalizes
naturally to any $\nu \geq 2$~\cite{gen_nu}.) In all cases, the
fractionalization is temporary, and after a sufficiently long
time, a charge unity object is recovered. However, the fractional
charges can be measured directly by using time resolved or finite
frequency
measurements. %A direct way to measure fractional charges in LL's using a
%zero frequency method is yet to be found.

\emph{Acknowledgement} We thank M. Heiblum, A. Yacoby, B. I.
Halperin, I. Neder, J. Eisenstein, K. Le Hur and S. Kivelson for
valuable discussion. The research was supported by the grants
DIP-H.2.1, DOE\# DE-FG02-06ER46287 (EB), DAAD (FvO) and the
Stanford Institute of Theoretical Physics (EAK).


\begin{thebibliography}{99}
\bibitem{comment-tunneling} Inter-edge tunneling is relevant in the renormalization group sense.
However, as long as we are at not too low temperatures (or bias voltages),
its effect can be small.
\bibitem{Comment08-smooth-transition} The total reflected charge in each
scattering event does not depend on the exact shape of the
transition region.
\bibitem{gen_nu} In the $\nu\ge2$ case, the reflected charge is $\frac{ne}{%
\nu}$, where $n$ is the number of modes which are reflected by the point
contact.
\end{thebibliography}
\end{document}